# A polarity-induced defect mechanism for conductivity and magnetism at polar-nonpolar oxide interfaces

Liping Yu and Alex Zunger
University of Colorado, Boulder, Colorado 80309, USA

**The discovery of conducting two-dimensional electron gas (2DEG) and magnetism at the interface between insulating nonmagnetic oxides, as exemplified by the polar LaAlO$_3$ and nonpolar SrTiO$_3$ has raised prospects for attaining interfacial functionalities absent in the component materials. Yet, the microscopic origin of such emergent phenomena remains unclear, posing obstacles to design of improved functionalities. Using first principles defect calculations, we reveal a unifying polarity-induced defect mechanism for both conductivity and magnetism at polar-nonpolar interfaces of nonmagnetic insulating oxides. We demonstrate that the polar-discontinuity across the interface triggers thermodynamically the spontaneous formation of certain defects that in turn cancel the polar field induced by the polar discontinuity. It turns out that the 2DEG originates from those spontaneously formed surface donor defects (oxygen vacancy), but the density of 2DEG is controlled by the interfacial anti-site acceptor defects (Al-on-Ti). The interface magnetism is found to originate from the un-ionized deep Ti-on-Al anti-site donor defects within the LaAlO$_3$ side near the interface. Our results suggest practical design guidelines for inducing and controlling both 2DEG and magnetism at polar-nonpolar oxide interfaces.**

Oxide interfaces exhibit many spectacular phenomena not found in the respective bulk components or in conventional semiconductor interfaces[1], providing new avenues for electronics[2]. The LaAlO$_3$/SrTiO$_3$ interface is a paradigm example, exhibiting conducting 2DEG[3,4] and magnetism[5-11] between two insulating nonmagnetic metal-oxides. In the [001] direction, two different interfaces can be formed between polar LaAlO$_3$, which consists of alternating (LaO)$^+$-(AlO$_2$)$^-$ layers, and nonpolar SrTiO$_3$, which consists of alternating (SrO)$^0$-(TiO$_2$)$^0$ layers. One is called n-type (i.e., LaO/TiO$_2$) and the other is called p-type (i.e., AlO$_2$/SrO). The remarkable feature is that the conductivity occurs only at n-type interfaces when the LaAlO$_3$ film thickness ($n_{LAO}$) is larger than three unit cells (uc)[4,5], whereas the magnetism has been observed both at n-type interfaces with $n_{LAO} >$ ~3 uc and at insulating p-type interfaces[8]. As listed in Table 1, this feature and other experimental observations represent the main puzzles[12] that need to be resolved before the promised applications can be realized[13].

*Existing explanations:* At present, no single mechanism can fully explain the vast and growing body of experimental work on this LaAlO$_3$/SrTiO$_3$ system. *For 2DEG at n-type interfaces*, four main mechanisms have been suggested. The prevalent one is intrinsic electronic reconstruction (so called polar catastrophe) involving ionization of the host valence band of LaAlO$_3$ within the abrupt and defect-free interfaces[3,4]. The other three mechanisms involve various defects, including the oxygen vacancies at the interface (denoted as V$_O$(*I*), where "*I*" means *Interface*)[14-16], oxygen vacancies at LaAlO$_3$ overlayer surface (denoted as V$_O$(*S*), where *S* means *Surface*)[17-22], and the La-on-Sr (La$_{Sr}$) antisite donor defects induced by interfacial cation intermixing[23-29]. As Table 1 shows, each mechanism represents one aspect of the interface physics, explains some experimental findings, but conflicts with a few others[2]. None explains the insulating nature of p-type interfaces. *For interface magnetism,* it was shown experimentally that the local magnetic moments are associated with Ti$^{3+}$ ions[5-11,30]. However, it is yet unclear whether such Ti$^{3+}$ ions reside in SrTiO$_3$, or LaAlO$_3$, or both near the interface. Theoretically, it has been postulated that Ti$^{3+}$ ions arise in SrTiO$_3$ side, involving the occupation of the low-energy Ti-$d_{xy}$-like sub-bands caused by the interfacial splitting of orbital degeneracy[31], or interfacial disorder[32,33], or interfacial oxygen vacancies[34], which are difficult to reconcile why the magnetism appears at insulating p-type interfaces and at n-type interfaces with a critical thickness ($L_c$) similar to that for 2DEG.

***Overarching unresolved questions:*** a crucial issue associated with the emergent conductivity and magnetism at polar-nonpolar interfaces is what mitigates the divergence of electrostatic potential as the thickness of the polar film increases[35]. Is it electronic reconstruction within polar catastrophe scenario, atomic reconstruction, or chemical defects? Different answer represents different mechanism, suggesting different controlling parameters. It has been purported that all these three effects exist to varying degrees in real materials and contribute to the emergent interface phenomena. However, whether and how these effects are physically connected, and what are their relative contributions, remain uncertain. Particularly, for defects, it is unclear which specific defects can be induced and are responsible for the emergent interface phenomena. Using first principles electronic and defect calculations, we identify the specific defects that can form spontaneously as a response to the built-in polar field and show how these defects lead to a unifying mechanism (**Figure 1**) that simultaneously explains the main features of both conductivity and magnetism at the interface.

***The key physical quantities*** that feature in our explanation are (i) the formation energy $\Delta H$ (which controls the defect concentration) (**Figure 2**) of the defect in various charge states at thermodynamic equilibrium Fermi-energy $E_F$, and (ii) the donor or acceptor electrical levels (**Figure 3**), i.e., the defect charge transition energy $\varepsilon(q/q')$ defined as the $E_F$ where the $\Delta H$ of a defect at two different charge states $q$ and $q'$ equal. A donor can produce electrons and compensate holes, whereas an acceptor can produce holes and compensate electrons. These two quantities (i) and (ii) have not been previously calculated for charged defects in different layers across the interfaces and turn out to be crucial. The details of their first principles calculations are given in Methods section.

***The central point of the proposed mechanism*** is that the polar-discontinuity induced built-in polar field triggers thermodynamically the spontaneous formation of the certain defects at surface and/or interface, which in turn compensate the built-in polar field and thus avoids the potential divergence. Thus, it is the polar-field induced defects, rather than the electronic or atomic reconstruction, that are responsible for the conductivity and magnetism at the interface. Specifically, we find that the surface donor defect (here $V_O$) has its donor levels located energetically above the $SrTiO_3$ conduction band at the interface but below the $LaAlO_3$ conduction band. This donor level position is a prerequisite for 2DEG formation. Although the 2DEG owes its *existence* to the surface donors, the *density* of 2DEG is controlled by the interfacial deep acceptor defects (mainly Al-on-Ti antisite). It is also turn out that the interface magnetic moment is caused by the unionized deep Ti-on-Al antisite defects located within the $LaAlO_3$ side near the interface.

In what follows we address specifically how this polar-field induced defect mechanism resolves the long-standing puzzles on the origin of 2DEG, the critical thickness for 2DEG, the weak field in $LaAlO_3$ film, the density of 2DEG, the insulating p-type interfaces, and the origin of the local magnetism moments. During this process, we also distill the general rules that control the pertinent effects and allow future section of other polar-nonpolar interface materials with similar or improved interface properties.

**1. What creates the 2DEG?**

*The 2DEG is unlikely to originate from the defect-free scenarios:* these include the ionization of the intrinsic LaAlO$_3$ valence bands (suggested by the polar catastrophe model[3,4]) or the ionization of the LaO interface layer (suggested by the interfacial charge leaking model)[36] (Supplementary note 1). This is simply because that the creation of 2DEG in these defect-free scenarios requires the LaAlO$_3$ valence band maximum (VBM) to cross the SrTiO$_3$ conduction band minimum (CBM) or Fermi-energy $E_F$. This is contrary to the experimentally observed weak field (negligible band-bending)[37-40] in the LaAlO$_3$ film, clearly showing that the LaAlO$_3$ VBM is far below the $E_F$.

*The 2DEG also is unlikely to originate from interfacial point donor defects* (La$_{Sr}$, Ti$_{Al}$, and V$_O$). Recall first that the defect formation energy $\Delta H$ depends on the $E_F$ (or chemical potential) and the defect charge transition energy $\varepsilon(q/q')$ needs to be close to band edges in order to producing free carriers. In thermodynamic equilibrium, the $E_F$ of the system pins around the middle of SrTiO$_3$ band gap when $n_{LAO} < L_c$, and around the SrTiO$_3$ conduction band edge near the interface when $n_{LAO} \geq L_c$ (Supplementary note 2). In either case, Fig.2ab shows that the $\Delta H$ of the anti-site donor defects, La$_{Sr}^0$ and Ti$_{Al}^0$ is small positive or even negative (note: the superscript denotes the *defect charge states*, not the nominal *oxidation state* of the atom at the defect site). In other words, the formation of such antisite defects at the thermodynamic equilibrium $E_F$ is energetically favorable and would inevitably lead to interfacial cation mixing. However, at such $E_F$, both La$_{Sr}^0$ and Ti$_{Al}^0$ defects are stable in their charge *neutral* states (as indicated by the superscript), contributing no free carriers. On the other hand, the interfacial V$_O$ defects are energetically stable in the charged states, i.e., V$_O^{2+}$ (Fig.2ab). This means that if formed, the V$_O$ will donate electrons and thereby become positively charged. However, the $\Delta H$ of V$_O^{2+}$ at such equilibrium $E_F$ is rather high (> 2.5eV), implying that V$_O^{2+}$ have very low concentration under thermodynamic equilibrium conditions. The high $\Delta H$ also means that even if the V$_O$ defects are formed under non-equilibrium growth conditions, they can still be removed easily by the post O-rich annealing process[41] (Supplementary note 3). Thus, contrary to earlier postulations, these interfacial donor defects are not responsible for 2DEG, consistent with recent experiments[42].

*The oxygen vacancy, V$_O$(S), at the free LaAlO$_3$ surface can explain the interfacial 2DEG.* For this to happen, three conditions ("design principles") need to be satisfied. First, V$_O$(S) in the polar film material needs to have a sufficiently low formation energy $\Delta H$ so it could form. Fig.4a shows the $\Delta H$ of V$_O$(S) decreases linearly as the film thickness $n_{LAO}$ increases, consistent with previous calculations[20,43]. When $n_{LAO} \geq$ 3-4 uc, the $\Delta H$ becomes zero or negative, and V$_O$(S) will form spontaneously and be stable even for high oxygen chemical potentials. (This suggests that even exposing the surface to air or post-annealing under O-rich environment cannot heal these vacancies). Second, the donor transition level of V$_O$(S) in the polar film material should be higher in energy than the substrate (SrTiO$_3$) conduction band edge at the interface. Fig.3 shows that this condition is also satisfied. Third, the system needs to have a none-zero built-in polar field that would enable the electron to transfer from the surface to the interface. Such transfer sets up an opposite dipole (proportional to $n_{LAO}$), which in turn cancels the field and lowers the $\Delta H$. The larger the $n_{LAO}$, the lower the $\Delta H$. Note that in the absence of such field, the surface-to-interface charge transfer would not occur since such a transfer would create a dipole that would increase the electrostatic energy (proportional to that dipole) and thus raise the total energy of the system. These three conditions are satisfied in this LaAlO$_3$/SrTiO$_3$ system.

It is important to note that the built-in polar field always exists during the layer-by-layer growth. This is because that the surface defects (here V$_O$) can cancel the built-in polar field only in the LaAlO$_3$ film between the interface and the surface, not the built-in polar field in the LaAlO$_3$ film to be grown on top of the surface. Such polar field can then always trigger V$_O$ formation at the surface of LaAlO$_3$ film to be grown, no matter whether the polar field has been cancelled in the LaAlO$_3$ substrate or not.

***The emerging picture for the formation of the 2DEG*** is that the electrons ionized from V$_O$(*S*) of the polar film material transfer to the nonpolar substrate material SrTiO$_3$ conduction bands at the interface via the built-in polar field, thus forming the 2DEG at that interface. This charge transfer in turn cancels the built-in polar field in LaAlO$_3$. After the built-in field has been cancelled, the Δ*H* of V$_O$(*S*) return to a high value (> 3 eV) characteristic of the bulk, and V$_O$(*S*) become again hard to form in thermodynamic equilibrium[20]. Thus, the maximum concentration of V$_O$(*S*) is 0.25/$S_{2D}$ (where $S_{2D}$ is 2D unit cell area), i.e., one of eight oxygen missing at surface, donating 0.5 e/$S_{2D}$ that cancels the polar field in LaAlO$_3$ completely. The compensation of polar field by V$_O$(*S*) also means that the band bending in LaAlO$_3$ due to polar field is also removed. Thus, the LaAlO$_3$ valence bands fall well below the $E_F$, contrary to what the polar catastrophe model would suggest. Consequently, no free holes can arise from depopulation of the LaAlO$_3$ valence bands at the surface, consistent with experiments[3,22].

***The emerging design principles for selecting materials that will likely form interface 2DEG:*** (i) the nonpolar material has a conduction band minimum (CBM) sitting in the band gap of the polar material; (ii) The polar material has at least one *deep* donor defect with its donor level higher than the conduction band of the nonpolar material at the interface. The (i) and (ii) in together guarantee that the 2DEG forms only at the n-type interfaces. This picture suggests that the 2DEG at n-type LaAlO$_3$/SrTiO$_3$ interfaces may also be induced and/or tuned by using some surface adsorbates (e.g., H$_2$O, H)[44-46] or metallic contacts[47] provided that the ionization energy of the surface adsorbate or the metallic contact is not lower than the donor level of the V$_O$(*S*).

## 2. What controls the critical thickness?

The linear decrease of *ΔH* of V$_O$(S) with increasing polar film thickness $n_{LAO}$ naturally explains the critical thickness L$_c$ for the metal-insulator transition. The rate of decrease (i.e., the slope dΔ*H*/d$n_{LAO}$) equals 0.19 eV/Å, which is same as the calculated built-in polar field in the defect-free LaAlO$_3$ film (Supplementary note 4). The V$_O$(*S*) defects start to form spontaneously when the Δ*H* becomes zero at an $L_c$ of ~4 uc under a typical O-rich growth condition (Fig.3a). For the LaAlO$_3$ film that is one unit cell thinner than this $L_c$, the calculated Δ*H* of V$_O$(*S*) is 0.75 eV, which is too high to produce significant free carrier concentration. Thus, *the appearance of V$_O$(S) (and the ensuing metal-insulator transition) at L$_c$ is predicted to be sharp, distinct from the gradual appearance of 2DEG behavior as predicted from polar catastrophe model, but consistent with experiments*[48].

Fig.3a suggests that the $L_c$ resulting from V$_O$(*S*) is controlled by the formation energy Δ$H_o$ of V$_O$ at interface (or the Δ*H* extrapolated at $n_{LAO}$ =0) and the polar field $E_p$ via $L_c = ΔH_o/eE_p$. Since $E_p=4πP_0/ϵ$ (where $ϵ$ and $P_0$ are the dielectric constant and formal polarization of LaAlO$_3$ film), this relation can be written as $L_c = ΔH_o ϵ/4πeP_0$, which predicts an $L_c$ of ~4 uc, depending slightly on the O-poor/rich growth conditions (Supplementary note 5). The relation of $L_c = ΔH_o ϵ/4πeP_0$ provides an alternative explanation for the observed variation of the $L_c$ with the fraction x in (LaAlO$_3$)$_{1-x}$(SrTiO$_3$)$_x$ overlayer (where $P_0$ is proportional to x)[49]. This observation was originally explained by $L_c = ΔΦϵ/4πeP_0$ (where ΔΦ is the energy difference between LaAlO$_3$ VBM and SrTiO$_3$ CBM) within polar catastrophe model[49]. Since ΔΦ and Δ$H_o$ have similar value (~3-4 eV), it is not surprising that the $L_c$ predicted from these two models is also similar.

***Implication on the design of carrier mobility:*** **(i)** The relatively high 2DEG mobility may be enabled by a modulated doping effect[50], whereby the source of carriers (here at the LaAlO$_3$ surface) is spatially separated from the place where the carriers reside (here at the LaAlO$_3$/SrTiO$_3$ interface), thus minimizing carrier scattering by the defects. This minimal spatial separation is measured by the critical thickness $L_c$. The relationship $L_c = \Delta H_o \epsilon / 4\pi e P_0$ suggests that *a large $L_c$ (hence maintaining good mobility) could be achieved by selecting a polar materials with small polarization, large dielectric constant, and donor defects characterized by high formation energy $\Delta H$ at the interface or in the bulk.* On the other hand, **(ii)** the concentration of interfacial defects should be minimized in order to take advantage of (i). In addition, **(iii)** since the 2DEG is located at the conduction bands of the nonpolar material, it is advantageous to select the nonpolar material with low electron effective mass in order to achiever higher mobility.

### 3. What compensates the built-in polar field?

Experimentally, only very weak residual field has been observed in the LaAlO$_3$ film no matter whether its thickness is below or above the $L_c$[37-40]. This observation cannot be explained within the defect-free interface scenario. In turn, whereas the V$_O$(S) model explains the weak electric field in LaAlO$_3$ film *above* the $L_c$, it does not explain it *below* the $L_c$. This leads us to inspect the effects of all possible cation antisite defects across the interface.

***In term of point defect, each interfacial antisite alone cannot cancel the polar field.*** Fig.2ab shows that the La$_{Sr}$, Sr$_{La}$, Ti$_{Al}$, and Al$_{Ti}$ antisite defects have lower $\Delta H$ than other point defects (e.g., vacancy) in the layer where they are located. This means that these four antisite defects are the dominating defects in their corresponding layers. Apparently, the interfacial La$_{Sr}$ donor in SrTiO$_3$ side cannot set up an opposite dipole across the LaAlO$_3$ film that can cancel the polar field inside the LaAlO$_3$ film. For the Ti$_{Al}$ donor in LaAlO$_3$ side, since its donor level is lower than the SrTiO$_3$ conduction band at the interface, the ionized electrons cannot be transferred to the latter either to cancel the polar field. For interfacial Al$_{Ti}$ and Sr$_{La}$ acceptors, the polar field compensation is similar to that in polar catastrophe model: before the LaAlO$_3$ VBM reaches the acceptor levels of Al$_{Ti}$ or Sr$_{La}$, the polar field cannot be cancelled; After that, the electrons start to transfer from LaAlO$_3$ valence bands to the acceptor levels of these defects, and the internal field decreases gradually as $n_{LAO}$ increase, approaching zero at infinitely large $n_{LAO}$.

***The [Ti$_{Al}$+Al$_{Ti}$] defect pair is the most potent source of polar field cancellation among those donor-acceptor antisite defect pairs at n-type interfaces.*** The four leading antisite defects can form four types of donor–acceptor pairs: [Ti$_{Al}$+Al$_{Ti}$], [La$_{Sr}$+Sr$_{La}$], [La$_{Sr}$+Al$_{Ti}$], and [Ti$_{Al}$+Sr$_{La}$], which are denoted as ①,②,③,④ respectively in Fig.3. Clearly, the electron-transfer from donor to acceptor in both pair ② and ③ is unlikely since it will create a dipole in the *same direction* as the intrinsic dipole in LaAlO$_3$, and thus *increase* the dipole moment (also the electrostatic energy) and destabilize the interface. In both pair ① and ④, the charge transfer can cancel the polar field. However, the electron transfer in pair ① is energetically much more favorable because (i) Al$_{Ti}$ has a lower acceptor level than Sr$_{La}$ and (ii) the donor-acceptor separation distance (also the associated opposite dipole moment that lowers the total energy of the system) is larger in pair ① (Fig.3a).

***For $n_{LAO} < L_c$, the [Al$_{Ti}$+Ti$_{Al}$] can form spontaneously via Ti$\Leftrightarrow$Al exchange across the interface and cancel the polar field.*** Fig.4b (filled symbols) shows that the energy required to form such defect pair is negative (i.e., exothermic), and the largest energy gain is obtained when a Ti atom of TiO$_2$-interface monolayer is exchanged with an Al of AlO$_2$-surface monolayer, i.e., Al$_{Ti}$(*I*)+Ti$_{Al}$(*S*), which is consistent with previous first-principles calculations[51]. This means that Ti atom at the interface would hop to the AlO$_2$ surface layer and exchange with Al atom there. Similar to that of V$_O$(*S*), the linear decrease of $\Delta H$ with increasing donor-acceptor separating distance (Fig.4b) is a sign of polar field compensation by electron transfer from Ti$_{Al}$ donor to Al$_{Ti}$ acceptor, which is expected since the donor level is higher than acceptor level (Fig.3a). Fig.4a also shows that the V$_O$(*S*) has too high $\Delta H$ to form for $n_{LAO} < L_c$ (Fig.4a). Therefore, the polar field is cancelled by those spontaneously formed [Al$_{Ti}$(*I*)+Ti$_{Al}$(*S*)] pairs. On the other hand, since these defects are deep, they cannot cause free carriers in the both interface and surface regions (whence insulating).

***For $n_{LAO} \geq L_c$, the polar field is cancelled by spontaneously formed V$_O$(S), not by [Al$_{Ti}$(I)+Ti$_{Al}$(S)].*** Recall that the polar field always exists in the LaAlO$_3$ layers during the lay-by-layer growth. The existence of this polar field can always trigger the formation of V$_O$(*S*) and/or Ti$_{Al}$(*S*) defects as $n_{LAO}$ increases. For $n_{LAO} \geq L_c$, both V$_O$(*S*) (Fig.4a) and [Al$_{Ti}$(*I*)+Ti$_{Al}$(*S*)] pair (Fig.4b) have zero or negative $\Delta H$, meaning that both could form in ideal interfaces. However, if both V$_O$(*S*) and Ti$_{Al}$(*S*) present, since V$_O$ has a higher donor level than Ti$_{Al}$ (Fig.3a), the electron transfer from V$_O$(*S*) to Al$_{Ti}$(*I*) is energetically more favorable than that from Ti$_{Al}$(*S*) to Al$_{Ti}$(*I*), and the polar field is thus always cancelled by the former. After the polar field has been cancelled by V$_O$(*S*), Fig.4b (open symbols) shows that the $\Delta H$ of [Al$_{Ti}$(*I*)+Ti$_{Al}$(*S*)] pair becomes positive (0.4-0.7 eV), meaning that [Ti$_{Al}$+Al$_{Ti}$] pairs cannot be formed via Ti⇔Al exchange over a distance beyond $L_c$. Therefore, for $n_{LAO} \geq L_c$, only V$_O$(*S*) defects can form and cancel the polar field.

## 4. What controls the density of 2DEG?

***Reinterpretation of the puzzle:*** According to Gauss' law, the experimentally observed weak electric field in LaAlO$_3$ film means that the total external charge density (mobile and/or immobile) at the interface must be ~0.5 e/$S_{2D}$ (Supplementary note 6). For $n_{LAO} < L_c$, not conductivity at interfaces means that all (or almost all) interfacial charge does not contribute to the conductivity. For $n_{LAO} \geq L_c$, only a fraction of 0.5 e/$S_{2D}$ interfacial charge is seen in transport and so the majority of 0.5 e/$S_{2D}$ charge does not contribute to the conductivity. The puzzle thus is why the ~0.5 e/$S_{2D}$ charge exists at the interface with any $n_{LAO}$ but only a small part of it contribute to conducting 2DEG when $n_{LAO} \geq L_c$. This puzzle cannot be explained within defect-free interface scenario by polar catastrophe model[3,4] or interfacial charge leaking model[36], since both predict zero interfacial charge for $n_{LAO} < L_c$ and an interfacial charge density much higher than the measured 2DEG density for $n_{LAO} \geq L_c$ (Supplementary Fig.S1). The possibility of multiple carrier types (i.e., the electrons occupying $d_{xy}$ and those occupying $d_{xz}/d_{yz}$ sub-bands contribute differently in transport) at ideal interfaces has also been suggested to explain the measured 2DEG density above the $L_c$[52-55]. Similarly, the existence of such multiple carrier types could not explain the 0.5 e/$S_{2D}$ interface charge that indeed exists. Moreover, it is also difficult to reconcile why a full carrier density of 0.5 e/$S_{2D}$ has been observed at GdTiO$_3$/SrTiO$_3$ interfaces (where the same multiple carrier types exist)[56].

***The 2DEG density is controlled by the concentration of immobile acceptor defects that can trap itinerant electrons.*** Within the emerging defect picture, the total interfacial charge is always ~ 0.5 e/$S_{2D}$, which corresponds to the (almost) complete polar field cancellation. In SrTiO$_3$ side (where the 2DEG is located), there are mainly three types of acceptor defects, namely, Al$_{Ti}$, V$_{Sr}$, and V$_{Ti}$. At equilibrium $E_F$, Fig.2ab shows that these acceptor defects all

prefer to stay in negative charge states, i.e., $Al_{Ti}^{1-}$, $V_{Sr}^{2-}$, and $V_{Ti}^{4-}$. (In other defect charge states, these defects have much higher $\Delta H$ and are not shown in Fig.2ab). It means that once these defects form they will trap free electrons from the system and get negatively charged. Among these acceptor defects, $Al_{Ti}^{1-}$ acceptors have the lowest $\Delta H$ and thus they the most potent electron-trapper. For $n_{LAO} < L_c$, the $Al_{Ti}$ defects resulting from Ti⇔Al exchange trap all free electrons transferred from $Ti_{Al}(S)$ defects, and hence no free carrier can occur. For $n_{LAO} \geq L_c$, due to $V_O(S)$, the $\Delta H$ of $[Ti_{Al}+Al_{Ti}]$ pair changes from negative to positive (Fig.4b), meaning that the concentration of $Al_{Ti}$ defect resulting from Ti⇔Al exchange is reduced, compare to that formed below the $L_c$. Therefore, the $Al_{Ti}$ defect concentration is not sufficient to trap all $0.5e/S_{2D}$ electrons transferred from $V_O(S)$. Therefore, only a small fraction of 0.5 $e/S_{2D}$ can contribute to interface 2DEG.

*The recently observed LaAlO₃ cation-stoichiometry effect on 2DEG formation[42] may also be understood within above picture.* For Al-rich LaAlO₃ film, where both A-site and B-site sublattices are fully occupied (hence having no cation vacancies), the $Al_{Ti}$ anti-sites are the only electron-trapping defects and the incomplete trapping of $0.5e/S_{2D}$ interface charge by $Al_{Ti}$ defects leads to interface conductivity. However, for La-rich LaAlO₃ film, where B-site sublattice is not fully occupied, the cation vacancies ($V_{Ti}$ and $V_{Al}$) also become main electron-trappers, in addition to $Al_{Ti}(I)$. Though, the concentration of $Al_{Ti}$ is reduced, each cation vacancy induced in La-rich film traps more electrons than an $Al_{Ti}$. The insulating character can be then attributed to the complete interfacial electron trapping by both interfacial cation vacancies and $Al_{Ti}(I)$.

*The picture of $Al_{Ti}(I)$ as main electron-trapping defects may be extended to SrTiO₃/GdTiO₃ interfaces.* The observed full carrier density of 0.5 $e/S_{2D}$ there[56] can be ascribed to the fact that both SrTiO₃ and GdTiO₃ have the same Ti atom at B-site sublattice and have no $Al_{Ti}$-like antisite defects at the interface.

*Implication on how to increase the density of 2DEG*: The above picture suggests that the main controlling factor for the interface carrier density is the concentration of the acceptor defects (mainly $Al_{Ti}$ in stoichiometric or Al-rich film), which should be minimized for enhancing carrier density. Such $Al_{Ti}$-like electron trapping defects may be completely removed by designing other oxide interfaces like GdTiO₃/SrTiO₃ interfaces, whose bulk components have a common cation atom with multiple valence states.

## 5. Why are the p-type interfaces insulating?

The intriguing fact is that the so-called p-type interfaces are not p-type (hole) conducting and are actually insulating. The defect-free polar-catastrophe model for p-type interface predicts a hole conducting interface and an electron conducting surface when $n_{LAO} > \sim 7.3$ uc (Supplementary Fig.1) in contradiction with the insulating behavior observed robustly in experiment. To explain this, defects thus be involved. The emerging defect picture below differs from the literature model based solely on interfacial hole-polaron[57] or interfacial hole-compensating $V_O$ defects[23,43], which assumes that the interface has holes arising from the depopulation of the intrinsic SrTiO₃ valence bands.

*Interfacial point defects can neither cause conductivity nor cancel the polar field.* Similar to that at n-type interfaces, the interfacial $La_{Sr}$ and $Ti_{Ar}$ are stable at their charge neutral states and have negligible or negative $\Delta H$ at equilibrium $E_F$. It means that they cause inevitable interfacial cation intermixing but induce no free carriers. The $V_O$ and other defects at the interface have too high $\Delta H$ to form and thus they do not produce free carriers either. Also for similar reason, each leading antisite defect ($La_{Sr}$, $Sr_{La}$, $Ti_{Al}$, or $Al_{Ti}$) alone at p-type interfaces cannot cancel the polar field.

*The spontaneously formed donor-acceptor defect pairs always cancel the polar field but do not induce free carriers.* In the p-type interfaces, the polar field points from surface to interface, which is opposite to that in the n-type interfaces. Among four donor-acceptor defect pairs as indicted in Fig.3b, the pair ② (i.e., [$La_{Sr}$+$Sr_{La}$]) is energetically most favorable in polar field cancellation. For $n_{LAO} < \sim$ 4uc, the [$La_{Sr}(I)$+$Sr_{La}(S)$] pairs have negative $\Delta H$ (Fig.4d) and can form spontaneously via La⇔Sr exchanges, whereas the [$La_{Sr}(I)$+$V_{La}(S)$] have too high $\Delta H$ to form (Fig.4c). So the polar field is cancelled by the charge transfer from $La_{Sr}(I)$ to $Sr_{La}(S)$, which can be expected from their relative defect levels (Fig. 3b) and their linear decreasing behavior in $\Delta H$ as a function of $n_{LAO}$ (Fig.4d). For $n_{LAO} \geq \sim$4uc, the $\Delta H$ of [$La_{Sr}(I)$+$V_{La}(S)$] become negative (Fig.4c) and can also form spontaneously. Since $V_{La}$ has a lower acceptor level than $Sr_{La}$ (Fig.3b), the polar field is cancelled by the charge transfer from $La_{Sr}(I)$ to $V_{La}(S)$, rather than to $Sr_{La}(S)$. In absence of electric field, Fig.4d (open symbols) indicates that the La⇔Sr exchanges cannot occur anymore over a distance of ~ 4uc. Unlike that in n-type interfaces, the $V_O(S)$ defects in p-type interface always have too high $\Delta H$ to form. The defects involved in polar field cancellation are all deep. The calculated equilibrium $E_F$ according to those point defects turns out to stay always around the middle of $SrTiO_3$ band gap. It means that both VBM and CBM are far away from the $E_F$ and there are no free carrier arising from the depopulation of VBM and CBM in both interface and surface regions (whence insulating).

*Implication on the design of two-dimensional hole conductivity:* Clearly, the formation of interfacial free holes is prevented by these spontaneously formed deep $La_{Sr}$ defects that have donor level higher than the VBM at the interface. So to induce interfacial hole conductivity, one should search for the polar-nonpolar interfaces where all such donors have too high enough formation energy to form or (ii) their donor levels below the VBM at the interface. Practically, the (ii) may be achieved more easily by searching for the polar material whose VBM is higher than the charge transition energy levels of those spontaneously formed interfacial donor defects.

## 6. What causes interface magnetism?

Distinct form previous models[31-34] that explain magnetism based on interfacial Ti(3+) *within the $SrTiO_3$ (i.e, not a defect)*, we find below that the local magnetic moment originates from the un-ionized deep $Ti_{Al}$ anti site defect (i.e., Ti(3+)-on-Al(3+) *within $LaAlO_3$* side near the interface. The interface magnetism depends on the concentration and spatial distribution of such $Ti_{Al}$ defects. This picture explains not only why the magnetism appears at n-type interfaces with a similar critical thickness to that for 2DEG, but also why the magnetism also appears at insulating p-type interfaces[8].

*What causes local moment:* For n-type interfaces, when $n_{LAO} < L_c$, the polar field in $LaAlO_3$ is cancelled by the charge transfer from $Ti_{Al}(S)$ defects to the interface. These formed $Ti_{Al}$ defects are thus ionized, i.e., $Ti_{Al}^{1+}$ (where superscript denotes the defect charge states). The Ti atom at this defect site has oxidation states of 4+, denoted as, Ti(4+) which has no local magnetic moment. When $n_{LAO} \geq L_c$, the polar field in $LaAlO_3$ is cancelled by the charge transfer to the interface from $V_O(S)$ instead of $Ti_{Al}$. All $Ti_{Al}(I)$ defects in $LaAlO_3$ film in absence of internal field are most stable in their charge neutral (or un-ionized) states, i.e., $Ti_{Al}^0$, where Ti appears as Ti(3+) oxidation state, having a finite local magnetic moment. Therefore, the interface magnetism at n-type interfaces due to those un-ionized $Ti_{Al}^0$ defects should also have a critical thickness of ~4uc. For p-type interfaces, it is the charge transfer among the defects other than Ti$_{Al}$ defects that cancels the polar field in in LaAlO$_3$. So all formed Ti$_{Al}$ defects there are not ionized, having local magnetic moments, and can cause interface magnetism.

***The magnitude of local moment:*** The local moment of a single Ti$_{Al}$ defect at the interface can be estimated from that in bulk LaAlO$_3$, which is 0.84$\mu_B$ from our hybrid functional calculation. For ferromagnetic order as observed in experiment, the total interface magnetic moment depends on the concentration of un-ionized Ti$_{Al}$ defects in LaAlO$_3$ and can be very small per Ti atom in average. The experimentally observed inhomogeneous landscape of magnetism that also varies from sample to sample[8,9] may be attributed to the various spatial distributions of Ti$_{Al}$ defects, which may be sensitive to sample preparation conditions (such as temperature and $P_{O2}$) and local strain.

***Comparing with $V_O(I)$ model:*** Relative to V$_O$(*I*) as the origin of local moment, the Ti$_{Al}$(*I*) defects within LaAlO side are more reasonable in at least two aspects. First, the deep Ti$_{Al}$ defect is spatially localized and has an unambiguous local moment. While for V$_O$, which is a shallow donor, it donates electrons to the lower-energy interfacial Ti $d_{xy}$ sub-bands that have light effective mass inside the interface plane[55], and the resulting Ti(3+) may then be itinerant. Second, the Ti$_{Al}$ defects would form easily or even inevitable due to its small or negative $\Delta H$, whereas the interfacial V$_O$ require significant energy to form and if formed they may be removed completely after annealing.

## Discussion

This work establishes a physical link between polar discontinuity and defect formation: the polar discontinuity triggers spontaneous formation of certain defects that in turn cancel the polar field induced by polar discontinuity. This link reveals a unifying mechanism that simultaneously explains the experimental observations listed in Table 1. This mechanism leads to a set of design rules for both conductivity and magnetism at the LaAlO$_3$/SrTiO$_3$ interfaces and enables the design of other polar-nonpolar interfaces (not limited to oxides) with similar or improved interface properties by first principles defect calculations.

Having ruled out the electronic reconstruction, interfacial V$_O$, and interfacial cation intermixing mechanism as the possible origin of 2DEG in our calculations, we conclude that the 2DEG at n-type interfaces with $n_{LAO} \geq L_c$ originates from the spontaneously formed V$_O$(*S*) defects. This conclusion stems from the finding that the donor level of deep V$_O$ in in LaAlO$_3$ side is higher than the SrTiO$_3$ conduction band edge at the interface. This finding explains why the formation energy of V$_O$(*S*) decreases linearly as $n_{LAO}$ increases. The linear decreasing relation reveals a general formula and some new controlling parameters for the critical thickness of sharp metal-insulator transition in absence of the electric field in the polar LaAlO$_3$ film.

Instead of causing 2DEG, we find that the defects resulting from interfacial cation anti sites play key roles in other aspects. Specifically, (i) the cation intermixing cancels the polar field only below the critical thickness for both-type of interfaces. (ii) the interfacial Al$_{Ti}$ acceptor defects control the density of 2DEG. (iii) the un-ionized interfacial Ti$_{Al}$ defects within LaAlO$_3$ side cause the local magnetic moments. The (iii) suggests that all previous models on interface magnetism based on Ti(3+) ions in SrTiO$_3$ side should be revisited. Importantly, we find that the cation mixing is confined near the interface and does not extend in LaAlO$_3$ side over more than ~4 uc, above which the cation intermixing is prevented by spontaneous formation of some surface vacancy defects. This spatial confinement is critical for realizing nanoscale interface magnetism.

This mechanism provides three distinctive predictions to be tested in experiment as further validation. (i) For n-type interfaces, the AlO$_2$-surface layer is dominated by Ti$_{Al}$ defects when $n_{LAO} < L_c$ and by V$_O$ defect when $n_{LAO} \geq L_c$. (ii) For p-type interfaces, the LaO-surface layer is dominated by Sr$_{La}$ and V$_{La}$ defects, respectively, below and above an $L_c$ of ~4 uc. (iii) Ti$^{4+}$ and Ti$^{3+}$ signals exist in both sides of the interface. The appearance of the Ti$^{3+}$ signals should not be taken as a sign of conductivity. Whether the Ti$^{3+}$ signals detected by photoemission below the $L_c$[21,39,58,59] can be truly assigned to those Ti$^{3+}$ ions in SrTiO$_3$ side should be revisited carefully. How these Ti$_{Al}$ local moments are ordered (ferromagnetic, or antiferromagnetic, or else) and whether and how they interact with the itinerant 2DEG are still open questions that should be investigated further.

## Methods

All calculations were performed using density functional theory (DFT) and plane-wave projector augmented-wave (PAW)[60] method as implemented in the VASP code[61]. An energy cutoff of 400 eV was used. The Brillouin zone was sampled by 8 × 8 × 1 and 4 × 4 × 1 **k**-point mesh for 1 × 1 and 2 × 2 in-plane supercell respectively. The atomic forces were relaxed to be less than 0.03 eV/Å. The in-plane lattice constant was fixed to 3.943 Å (the relaxed lattice constant of SrTiO$_3$ by GGA[62]). In slab calculations, the 4 uc (~16 Å) vacuum layer was used and the dipole correction was always applied to remove artificial dipole interactions[63]. The results in Fig.2 and Fig.3 were obtained by using HSE hybrid functional[64] on the top of the GGA relaxed structures.

The formation energy of a defect (*D*) calculated from $\Delta H_D^q(E_F, \mu) = E_D^q - E_H + \sum_\alpha n_\alpha (\mu_\alpha^0 + \Delta\mu_\alpha) + q(E_v + E_F)$, where $E_D^q$ and $E_H$ are the total energies of a supercell with and without defect, respectively, and *D* being in charge state *q*. $n_\alpha$ is the number of atoms of specie *α* needed to create a defect. $E_F$ is the Fermi energy relative to VBM ($E_v$). $\Delta\mu_\alpha$ is the relative chemical potential of specie *α* with respect to its elemental solid (gas) ($\mu^0$). The equilibrium Fermi-energy was calculated self-consistently according to charge neutrality condition[65]. The chemical potentials relative to their elemental solid (or gas) phase are taken as variables and are bounded by the values that maintain a stable host compound and avoid formation of other competing phases in thermodynamic equilibrium (Supplementary Fig.S2). The details of theory and calculations can be found in Ref.[66].

## Acknowledgements

This work was supported by the US Department of Energy, Office of Basic Energy Sciences as part of an Energy Frontier Research Centers, under the award No. DE-AC36-08GO28308 to National Renewable Energy Laboratory (NREL). The computation was done by using capabilities of the NREL Computational Sciences Center supported by the U.S. DOE office of Energy Efficiency and Renewable Energy, under Contract No. DE-AC36-08GO28308.

## Author Contributions

L.Y. carried out the calculations, analyzed the results and wrote the paper. A.Z. initiated this study and contributed to the analysis of the results and the writing of the paper.

**Table 1: List of the main puzzles and robust experimental observations at LaAlO$_3$/SrTiO$_3$ interfaces.**

| Interface structure | Experimental Observations | Polar Catastrophe | Cation Mixing | V$_O$ at interface | V$_O$ at surface | Current mechanism |
|---|---|---|---|---|---|---|
| n-type | 1. Critical thickness ($L_c$) = 4 uc | ✔ | ✘ | ✘ | ? | ✔ |
| | 2. 2DEG density < 0.5 $e$/S | ✘ | ? | ✘ | ✘ | ✔ |
| | 3. Weak $E$ in LaAlO$_3$ for $n_{LAO}$ < $L_c$ | ✘ | ? | ✘ | ✘ | ✔ |
| | 4. Weak $E$ in LaAlO$_3$ for $n_{LAO}$ ≥ $L_c$ | ✘ | ✘ | ✘ | ✔ | ✔ |
| | 5. LaAlO$_3$ surface: insulating | ✘ | ? | ? | ✔ | ✔ |
| | 6. Interface: cation intermixed | ✘ | ✔ | ✘ | ✘ | ✔ |
| | 7. Interface **magnetism** | ✘ | ? | ✘ | ✘ | ✔ |
| p-type | 1. Interface: insulating | ✘ | ? | ✘ | ? | ✔ |
| | 2. LaAlO$_3$ surface: insulating | ✘ | ✘ | ? | ? | ✔ |
| | 3. Interface: cation mixed | ✘ | ✔ | ✘ | ✘ | ✔ |
| | 4. Interface **magnetism** | ✘ | ? | ? | ✘ | ✔ |

The symbol of '✔' and '✘' mean that the mechanism agrees or disagrees, respectively with the experimental observation. The '?' symbol denotes uncertainty.

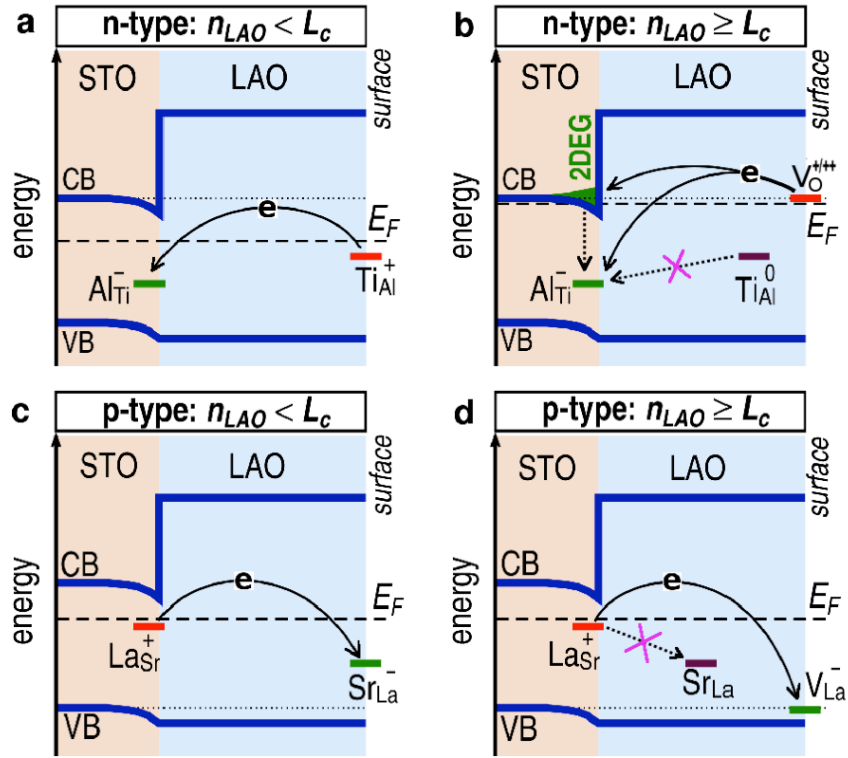

**Figure 1: Schematic band and defect level picture for the unifying mechanism. a,** n-type interfaces with $n_{LAO} < L_c$: The electrons transfer from $Ti_{Al}(S)$ to $Al_{Ti}(I)$ and compensate the electric field in $LaAlO_3$, inducing no itinerant carriers to the interface. **b,** n-type interfaces with $n_{LAO} \geq L_c$: The electrons transfer from $V_O(S)$ to interface. Part of interface charge is trapped by the deep $Al_{Ti}$ acceptor defects. The deep $Ti_{Al}^0$ donor defects are confined within $LaAlO_3$ near the interface and are not ionized, i.e., $Ti^{3+}$-on-$Al^{3+}$, having local magnetic moments. **c,** p-type interfaces with $n_{LAO} < L_c$ (~4uc): The electrons transfer from $La_{Sr}(I)$ to $Sr_{La}(S)$ and compensates the electric field in $LaAlO_3$. **d,** p-type interfaces with $n_{LAO} \geq L_c$: The electrons transfer from $La_{Sr}(I)$ to $V_{La}(S)$ and compensate the field in $LaAlO_3$. The superscripts (0,+,++,-) denote the defect charge states, not the oxidation states of the ions there.
12

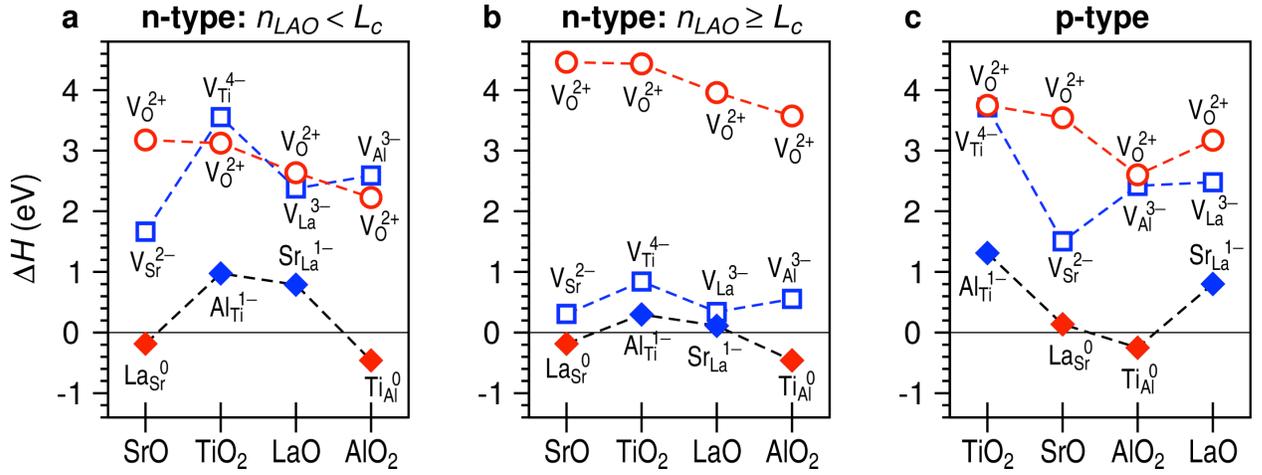

**Figure 2:** **Formation energy of the interfacial point defects at thermodynamical equilibrium Fermi energy. a,b,** n-type interfaces with $n_{LAO} < L_c$ and $n_{LAO} \geq L_c$, respectively. **c,** p-type interfaces. At a given $E_F$, the defect in different charge states (e.g., $V_{Sr}^0$, $V_{Sr}^{1-}$, $V_{Sr}^{2-}$) usually has different $\Delta H$ and the only one with the lowest $\Delta H$ is shown in the Figure. Other cation defects that have higher $\Delta H$ are shown in Supplementary Fig. 3. The chemical potentials used for Sr, Ti, La, Al, and O are -4.36, -6.20, -6.10, -5.46, and -2.0 eV respectively, relative to their corresponding elemental solid or gas phases, which corresponds to T=1050 K and $P_{O_2} = 6.1 \times 10^{-6}$ Torr.

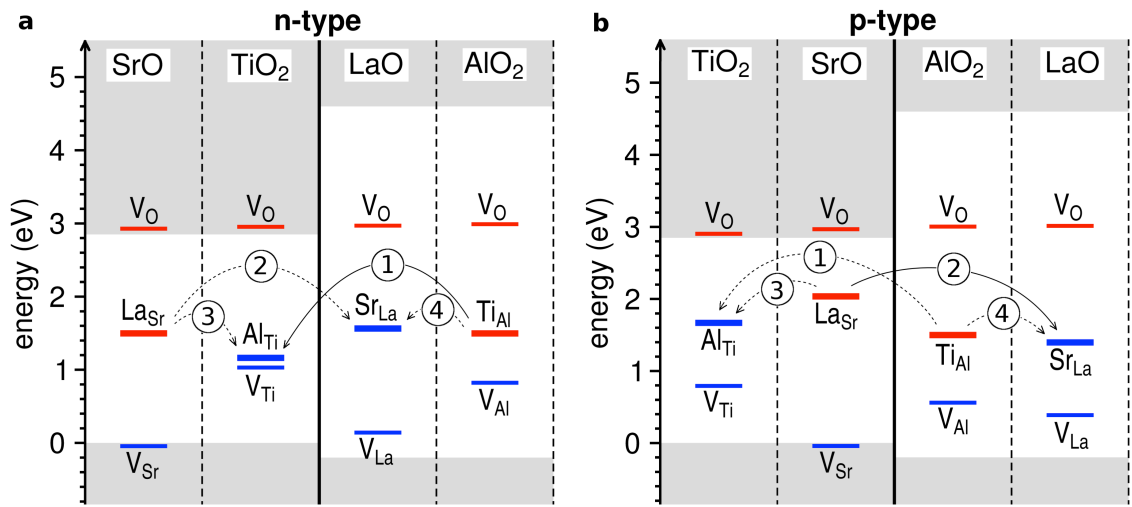

**Figure 3: Defect charge transition energy levels of the interfacial point. a,** n-type interface. **b,** p-type interface. The defect charge transition energy level is defined as the $E_F$ where the $\Delta H$ of a given defect in two different charge states equal. Some defects may have multiple charge transition energy levels. For example, $V_{Sr}$ has the two transition energy levels (one is for the transition between neutral charge state and -1, and the other is between -1 and -2). In such case, if the defect is donor (red), only the lowest level is shown, and if the defect is acceptor (blue), the highest level is shown.

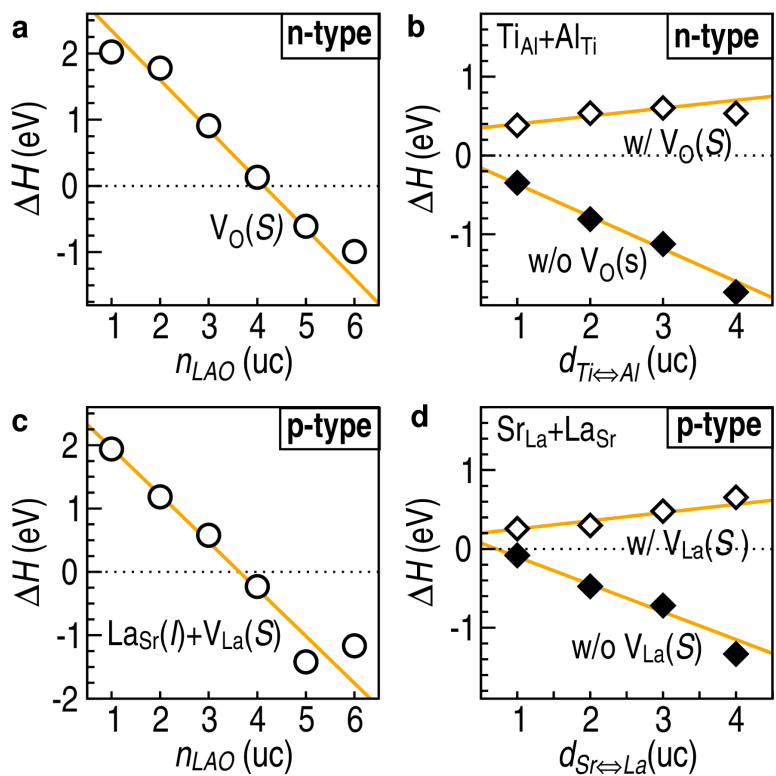

**Figure 4: Properties of surface defects and defect complexes. a,** the GGA-calculated $\Delta H$ of $V_O(s)$ defect, under the O-rich growth condition (i.e., $\Delta\mu_O = -1.5$ eV, Supplementary Fig. 2a). **b,** the $\Delta H$ of [$Ti_{Al}+Al_{Ti}$] defect pair created from a Ti⇔Al exchange out of the ideal interface with and without a $V_O(S)$ in a 2x2 6STO/4LAO/vacuum surpercell. **c,** the GGA-calculated $\Delta H$ of [$La_{Sr}(I)+V_{La}(S)$] defect complex as a function of $n_{LAO}$, under $\Delta\mu_{Sr} = -4.36$ eV (Supplementary Fig. 1b). **d,** the $\Delta H$ of [$La_{Sr}+Sr_{La}$] defect pair created from a La⇔Sr exchange out of the ideal interface with and without a $V_{La}(S)$ in a 2x2 6STO/4LAO/vacuum surpercell, respectively. The $d_{Ti\Leftrightarrow Al}$ and $d_{La\Leftrightarrow Sr}$ in **b,d** are the distance between the components of corresponding defect pair. The orange lines are the guides to the eye.

# Supplementary Information for "A polarity-induced defect mechanism for conductivity and magnetism at oxide interfaces" by Liping Yu and Alex Zunger

**Supplementary Note 1:**

**The failures of polar catastrophe model**

The key argument in favor of the polar catastrophe mechanism is its predicted critical thickness ($L_c$) of metal-insulator transition. Both electrostatics and the first principles calculations based on density functional theory (DFT) with the generalized gradient approximation (GGA)[1], shown in Fig. S1c lead to an $L_c$ between three and four unit cells[2,3]. However, these results may be clouded by specific uncertainties: the electrostatic model depends on the choice of the LAO film dielectric constant and LAO/STO band offsets, whereas the GGA calculation suffers from the well-known band gap underestimation problem[4]. To examine this point we have applied the HSE hybrid functional[5], which predicts correctly the experimental bulk STO band gap of 3.2 eV. We find an $L_c$ of 4-5 uc (Fig. S2c). It is noteworthy that, for $n_{LAO}$ = 4 uc, the band gap of the system turns out to be 1.1 eV, suggesting an insulating behavior, contrasting with the experiment where the robust conductivity has been observed at this thickness. Hence, strictly speaking, the defect-free polar catastrophe model cannot explain the $L_c$ either[6].

In a *defect-free* interface structure, both previous DFT-GGA calculations[7,8] and our HSE calculation (Fig.S2d) predict a large electric field of 0.19 V/Å in the LAO film of $n_{LAO} < L_c$, which decreases gradually as $n_{LAO}$ increases for $n_{LAO} \geq L_c$, approaching zero only at infinitely thick $n_{LAO}$. This prediction of the PC model conflicts with experiments, where only very weak residual field has been observed for all investigated $n_{LAO}$[9-12].

**Supplementary Note 2:**

**Thermodynamical equilibrium Fermi energy as a function of LaAlO$_3$ thickness**

Under a given growth condition, the thermodynamic equilibrium $E_F$ is established as a balance of all electrons and holes concentrations contributed by thermal ionization of all considered defects. According to charge neutrality condition[13], the $E_F$ can be calculated self-consistently from first principles using defect formation energies as inputs.

As seen in the main text, the charge transfer from surface defect to interface defect always cancels the built-in polar field. For the n-type interface below the critical thickness and p-type interfaces at any thickness, all involved defects are deep defects, the resulting equilibrium $E_F$ always sits near the middle of SrTiO$_3$ band gap. While for n-type interfaces above the critical thickness, the **surface** O vacancies (V$_O$) form spontaneously. This V$_O$ defect has a donor level higher than the SrTiO$_3$ conduction band at the interface. The electrons will be transferred from Vo donor level to the SrTiO$_3$ conduction bands and pins the E$_F$ near the SrTiO$_3$ valence band minimum.

**Supplementary Note 3:**

**Justification for the thermodynamic equilibrium**

The thin film growth is an out of equilibrium process. Our thermodynamic calculations may be still applied. The more detailed justification can be found in Ref [14]. In brief, the calculated formation energies and defect levels (deep or shallow) in this work are physically meaningful for both equilibrium and non-equilibrium processes. Non-equilibrium implies that once certain high-energy defects form, kinetic barriers may preserve them, even if their concentration exceeds the nominal equilibrium value. However, it should be clear that defects with high formation energy would always be unlikely to form, since a lot of energy needs to be expended in their creation, and the driving force to lowering the energy is large.

**Supplementary Note 4:**

**The formation $\Delta H$ of V$_O$(S) vs LAO film thickness $n_{LAO}$**

The linear decreasing of $\Delta H$ of V$_O$(S) with increasing $n_{LAO}$ can be understood easily from the opposite dipole created by the charge transfer from V$_O$(S) to the interface.

In a supercell calculation, the formation energy of an oxygen vacancy at LAO surface (denoted as V$_O$(s) hereafter) is

$$\Delta H = [\mathcal{E}_D^0 + en_{LAO}E'] - [\mathcal{E}_H^0 + en_{LAO}E] + \mu_O, \tag{1}$$

where $\mathcal{E}_D^0$ and $\mathcal{E}_H^0$ are the total energies of the supercell structures in the absence of an electric field across the LAO film with and without the vacancy. The second term in each of the brackets corresponds to the electrostatic energy rise due to the presence of internal electric field $E'$ and $E$ in LAO with and without V$_O$, respectively. In a 2×2 2D supercell considered here the creation of a single V$_O$(S) leads to zero electric field in LAO ($E'$= 0 for all $n_{LAO}$).[8,15,16] Thus, Eq. (1) reduces to $\Delta H = [\mathcal{E}_D^0 - \mathcal{E}_H^0 + \mu_O] - en_{LAO}E$, where the second term can now be viewed as an opposite dipole, which decreases linearly at a rate of $eE$ as $n_{LAO}$ increases. Therefore, the larger the $n_{LAO}$, the larger the opposite dipole moment that lowers the total energy of the system, and the more energy decrease in $\Delta H$. For V$_O$(I) defects, which can not create an opposite dipole that lowers the $\Delta H$, the $\Delta H$ is thus independent of $n_{LAO}$ and remains high as $n_{LAO}$ increases[8].

The slope d$\Delta H$/d$n_{LAO}$) remains same or almost same as V$_O$(S) defects accumulate gradually. This is because that the built-in polar field is created by the polar *charge* at surface/interface. Microscopically, the electron transfer from "one V$_O$(S)" to the "interface" mainly cancels the polar field caused by the "polar charge at this V$_O$(S) site", not the "polar field caused by the polar *charge* at other V$_O$(S) defect sites". It means that during gradual formation of V$_O$(S), though the in-plane-averaged electric field is lowered, each V$_O$(S) to be formed still face the (almost) same polar field to be cancelled by it. Hence, the decrease in $\Delta H$ of V$_O$(S) should be almost same as VO(S) defects accumulates. **This thus explains why the metal-insulator transition due to V$_O$(S) is sharp.**

**Supplementary Note 5:**

**Critical thickness vs oxygen chemical potential**

Eq.(1) suggests that the critical $L_c$ can be approximately determined by the $\Delta H$ of the interfacial V$_O$ in TiO$_2$ layer and the electric field E in the defect-free LAO film (from electrostatics), via

$$L_c = [\mathcal{E}_D^0 - \mathcal{E}_H^0 + \mu_O]/eE, \tag{2}$$

which is a function of the oxygen chemical potential $\mu_O$ and the built-in electric field in defect-free LAO. Supplementary Figure 1b shows the variation of resulting $L_c$ due to V$_O$(S) with respect to $\Delta \mu_O$ (i.e., the oxygen chemical potential relative to 1/2O$_2$) at $T$ = 0K. The corresponding $\Delta\mu_O$ of current growth conditions ranges from -1.73 to -2.25 eV, where the $L_c$ is found to be between 3.1 and 3.8 uc. The O-rich annealing conditions have $\Delta\mu_O$ ranging from -0.61 to -1.44 eV, where the resulting $L_c$ varies from 4.2 to 5.3 uc. Considering the possible high kinetic barrier associated with the low-temperature annealing, the O vacancies induced during the growth generally may not be completely compensated, and thus it is expected that the actually $L_c$ in the annealed samples be ~ 4 uc, consistent with experiments.

### Supplementary Note 6:

**Interfacial charge vs. internal electric field in polar film**

In absence of externally applied electric field, the electric displacement $D_{STO} = 0$ in SrTiO$_3$ side, and $D_{LAO} = E + 4\pi P$ in LaAlO$_3$ side. According to Gauss' law, the external interface charge $Q = D_{LAO} - D_{STO} = E + 4\pi P$. For $E = 0$, $Q = 4\pi P = 0.5$ e per two dimensional unit cell area, which is the maximum. For $E = -4\pi P$, $Q = 0$, which is the minimum, i.e., the case below the $L_c$ within the polar catastrophe model.

## Supplementary Figures

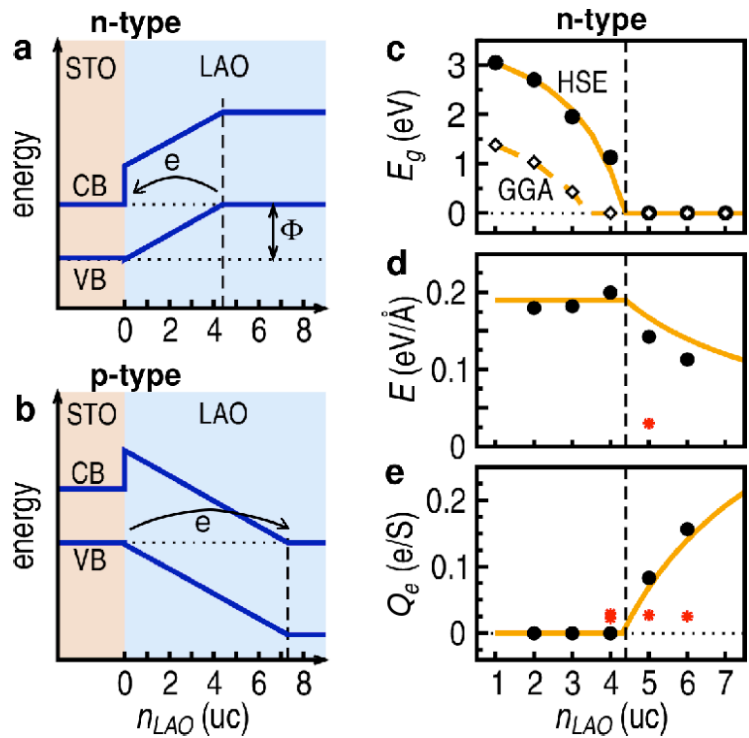

**Figure S1: Electronic properties of abrupt defect-free interfaces within the polar catastrophe scenario. a,b,** schematic band diagram for n-type and p-type interfaces respectively. **c**, GGA and HSE band gaps of 6STO/nLAO/vacuum interfaces. **d**, Calculated macroscopic planar-averaged electric field in the center of LAO film by HSE. The experimental data (*) is taken from ref.[10]. **e**, Calculated interfacial carrier density: HSE (black dots) vs. experiment (red stars)[17]. The orange lines in **d,e** are calculated from the simple electrostatic model using Φ=3.3 eV (see **a**) and the LAO film dielectric constant of 30, suggesting a critical

thickness of 4.3 uc for n-type interfaces. Using the same Φ and dielectric constant for p-type interfaces, the resulting critical thickness is 7.3 uc, which is shown in **b**.

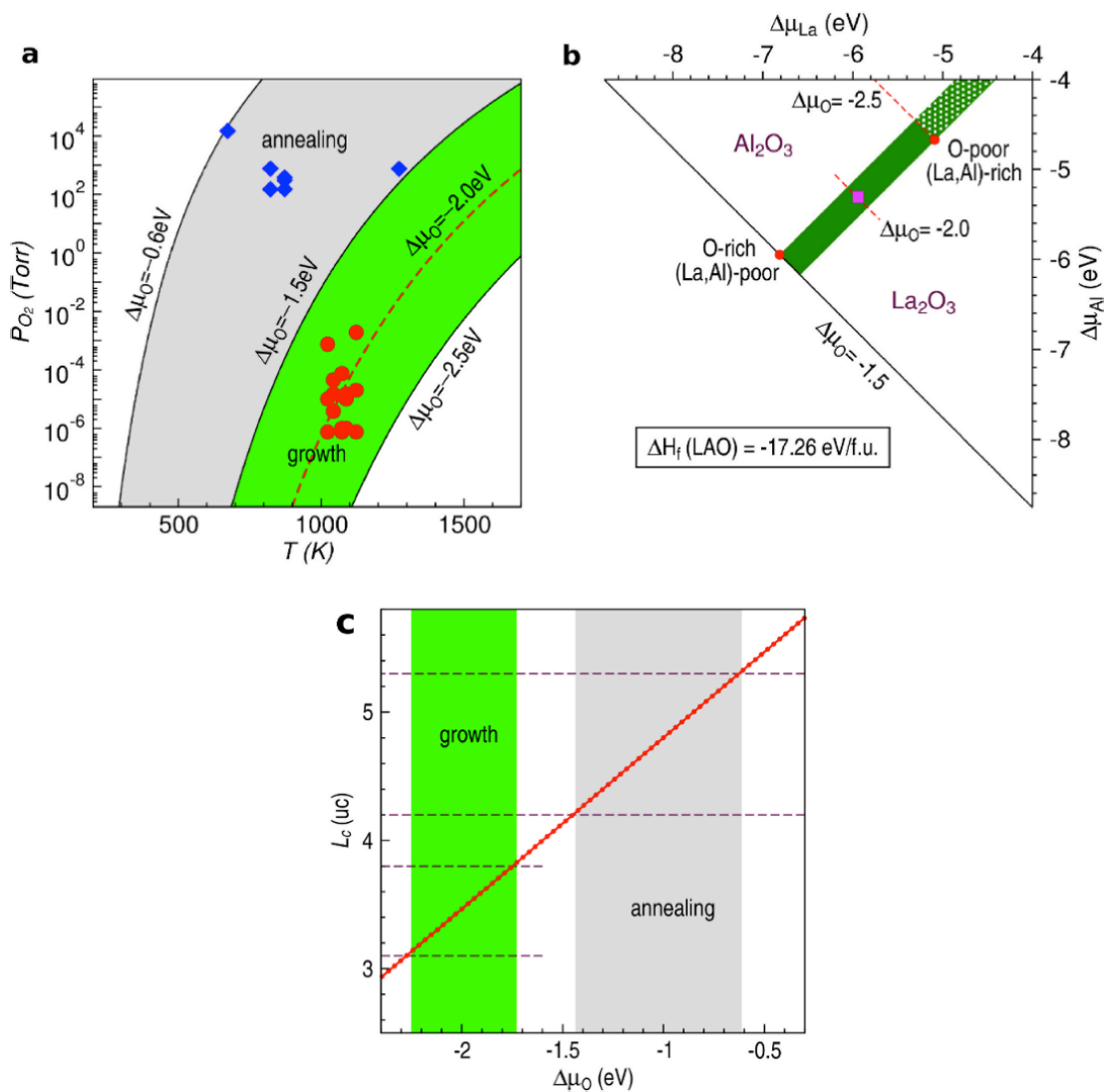

**Figure S2: Thermodynamic equilibrium growth conditions and critical thickness for STO/LAO interface.** **a**, typical growth (circles) and annealing (diamonds) conditions vs. oxygen chemical potential $\Delta\mu_O$ defined relative to $O_2$ molecule. The $\Delta\mu_O$ that corresponds to the experimental ($P_{O2}$, $T$) growth conditions was calculated according to the thermodynamic model[18]. The red dashed line corresponds to $T = 1050$ $K$ and $P_{O2} = 6.1\times10^{-6}$ Torr, which is referred in Fig.2. **b**, allowed equilibrium chemical potentials (green area) for growing LAO film on top of STO substrate, where the chemical potentials of elements satisfy $\Delta\mu_{La}+\Delta\mu_{Al}+\Delta\mu_O = \Delta H_f = -17.26$ eV and the completing phases ($Al_2TiO_5$, $SrAl_2O_4$, $La_2TiO_5$, $La_2Ti_2O_7$, $LaTiO_3$, $Al_2O_3$, $La_2O_3$, $SrO$, and $TiO_2$) cannot form. The experimental condition used for Fig.3 refer to $\Delta\mu_O = -2.0$ eV. **c,** the critical thickness ($L_c$) as a function of O chemical potential used during growth and annealing process.

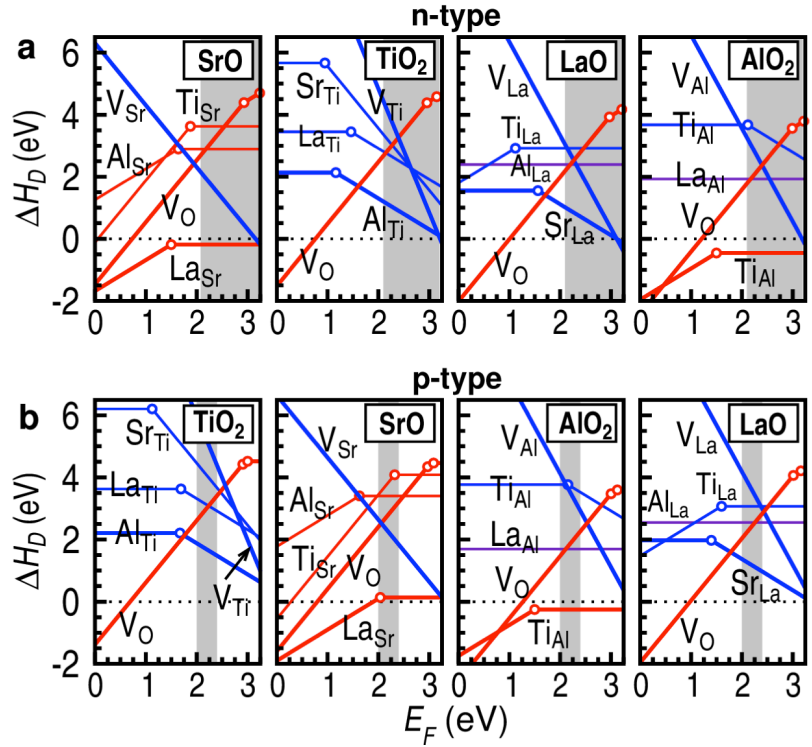

**Figure S3: Properties of interfacial defects in the 6STO/2LAO heterostructure containing both n-type and p-type interfaces. a,** n-type interface. **b,** p-type interface. Each panel in **a,b** shows various defects in a given atomic layer. Each line represents the $\Delta H$ of a donor (red) or acceptor (blue) defect. Different slops of line segments represent different charge states of a defect that are most stable at given $E_F$. Open circles mark the defect charge transition energies, i.e., the $E_F$ where the formation energy of a defect in two different charge states equal. The shaded regions in each panel denote the variation range of the equilibrium $E_F$. The chemical potentials for Sr, Ti, La, Al, and O are -4.36, -6.20, -6.10, -5.46, and -2.0 eV respectively, relative to their corresponding elemental solid or gas phases, which corresponds to T=1050 K and PO2 = $6.1 \times 10^{-6}$ Torr.